\documentclass[12pt]{article}
\usepackage{amsmath,cite}
\usepackage[dvips]{color}
\usepackage{hyperref}
\usepackage{shadow,amssymb,amsfonts,amsmath,fancybox,boxedminipage,times,epsfig,wrapfig}
\usepackage{psfrag,rotating}
\usepackage{graphicx}

\catcode`\@=11

\@addtoreset{equation}{section}

\long\def\@makefntext#1{\parindent 0cm\noindent
\hbox to 1em{\hss$^{\@thefnmark}$}#1}

\begin{document}
\begin{titlepage}
\vspace{.5in}
\begin{flushright}
UCD-04-27\\
gr-qc/0410015\\
October 2004\\
\end{flushright}
\vspace{.5in}
\begin{center}
{\Large \bf
Perturbation theory in covariant canonical quantization\\[1ex]
}
\vspace{.4in}
Sayandeb Basu\footnote{email:sbasu@lifshitz.ucdavis.edu}\\
\emph{\small{Department of Physics}}\\
\emph{\small{University of California}}\\
\emph{\small{Davis CA 95616}}\\
\emph{\small{USA}}
\end{center}
\vspace{.5in}
\begin{center}
{\large\bf Abstract}
\end{center}
\begin{center}
\begin{minipage}{4.75in}
{I investigate a new idea of perturbation theory in covariant canonical quantization. I present preliminary results for a toy model of a harmonic oscillator with a quartic perturbation, and show that this method reproduces the quantized spectrum of standard quantum theory. This result indicates that when the exact solutions to classical equations are not known, covariant canonical quantization via perturbation theory could be a viable approximation scheme for finding observables, and suggests a physically interesting way of extending the scope of covariant canonical quantization in quantum gravity}
\end{minipage}
\end{center}
\end{titlepage}

\addtocounter{footnote}{-1}

\section{introduction}
There is a long-standing debate in quantum gravity on the merits of covariant versus canonical approaches to quantization. Canonical approaches require a split of spacetime into space and time, contrary to the spirit of general covariance. covariant approaches based for instance on path integral formulation, is intrinsically perturbative and require a classical background. There is one approach, covariant canonical quantization, that avoids both these weaknesses, and provides a resolution to the covariant versus canonical debate, at least in principle. In this approach one quantizes the space of solutions to the classical equations. This solution space is isomorphic to the space of initial data (provided the initial value problem is well posed), and each initial datum is in turn a point on the phase space. Covariant phase space \cite{Ashtekar,Witten} provides a starting point based on observables of the theory, since canonical data are themselves observables in a time reparametrization invariant theory. In spite of this nice feature the method remains relatively unexplored, mainly because it relies on the knowledge of the full set of solutions to the equations of motion. To address this difficulty, it has been suggested \cite{SteveC} that it may be useful to extend the scope of this idea through perturbation theory. In this paper, I attempt such a construction. In particular I see if one can quantize a harmonic oscillator with a quartic perturbation by quantizing the space of solutions generated at each order via a classical perturbation expansion.

General relativity differs from other classical field theories in many aspects. The classical theory is nonlinear and its underlying symmetry, diffeomorphism invariance, implies that observables of the theory are non-local, a local definition of a field defined at a spacetime point $x^\mu$ is not physically meaningful, since diffemorphism actively moves one spacetime point to another. Diffeomorphism invariance also implies that there is no preferred background time. This raises problems for canonical quantization, which in its usual formulation requires a preferred splitting of spacetime into space and time. The absence of a preferred time makes the description of observables and dynamics one of the central challenges in quantum gravity \cite{SteveC, Isham, ku, ku2, sc, Bergmann, Torre}.  
In the canonical formulation, one works on a fixed time slice. Here we usually start with the canonical geometrodynamic variables $(q_{ab}, \pi^{ab})$ \cite{ADM} or the Ashtekar variables \cite{ROV}. The challenge then is to describe the dynamics of spacetime in an invariant way, after having chosen to work in the first place with an ``artificial'' split of spacetime into space and time. In classical general relativity the evolution from an initial time slice to a final one is independent of the choice of time in the intervening spacetime. But it is not clear if the same is true in the quantum theory---Torre and Varadarajan \cite{Torre3} recently showed that even in the absence of gravity, quantum evolution of a scalar field can be slice-dependent.

In the covariant picture, we work within a path-integral framework and employ a background field method that closely follows the standard approach to quantum field theories on a fixed background. Although the path-integral formally yields quantities that are covariant under diffeomorphisms of the background metric, the metric variables that we started are by themselves not diffeomorphism invariant. In contrast to the usual field theory approach, where correlators of locally defined fields make sense, in this case the correlators are complicated non-local functions of curvatures. If the perturbation $h_{\mu\nu}$ is large enough, the signature of the metric could change, and that will mean that the Lorentzian path-integral could miss the space of physical metrics. Further, Goroff and Sagnotti \cite{Goroff} have shown that gravity is nonrenormalizable at two loops --- a result that effectively ended hopes of a conventional quantum field theoretical approach to quantum gravity. Conventional perturbative evaluation of the path-integral thus seems untenable for quantum gravity. Hence, there is motivation to formulate a new approach to perturbation theory for quantum gravity.

In covariant canonical quantization, also known as covariant phase space quantization, the starting point is the space of classical solutions. For a theory with a well-posed initial value problem, the classical solutions are uniquely determined once arbitrary initial data is specified. Each set of initial data is also a point on phase space, and hence by restricting the classical solutions to an arbitrary initial value surface $\Sigma_{0}$, one achieves an isomorphism between the space of solutions $\mathbb{S}$ and the phase space $\textrm{T}^* \mathbb{Q}$ (see section 2 for details). The challenging aspect here is to find interesting physical observables---since the canonical data $(q, p)$ are time independent, the wave functions $\psi(q)$ or $\phi(p)$ in covariant canonical quantization are Heisenberg picture wave functions, and nothing seems to evolve. However, Rovelli \cite{Rovelli2, rovelli3, Rovelli3} has shown that the observables in this framework are ``evolving constants''--- one parameter families of phase space functions with vanishing total ``time'' derivatives. As we discuss in section 2, the canonical data are themselves  ``evolving constants'' in the sense defined by Rovelli. Time evolution is now encoded as a mapping between data. As noted above, the main obstacle in making progress with this idea (except in a few cases such as Carlip (2+1) \cite{Carlip, scj} and Torre \cite{cgtorre}(cylindrical waves)) is that in general relativity in 3+1 dimensions we do not know the exact general solutions, hence motivating a perturbative approach to extend the scope of covariant canonical quantization. 

The proposal of the present paper can be outlined in the following way. I employ a classical perturbation expansion around a known solution to construct the canonical data of a different problem that cannot be solved exactly.  Formally, this amounts to solving the Hamilton-Jacobi equations at each order, and then, in the spirit of the covariant canonical approach, identify the solutions at each order with canonical data of the full problem at that order. We can then pass onto a quantum theory by representing the perturbed canonical data as self-adjoint operators on the Hilbert space of basis states of the exactly soluble problem. 

Constructing observables by  such ``systematic approximations'' is not altogether a new idea. It was first proposed in the context of general relativity by Bergmann et al \cite{PGN}. Since then, the idea have been reiterated as an interesting alternative approach to the problem of observables by Carlip \cite{SteveC}, Torre \cite{Torre} and in a different guise by Gambini and Pullin \cite{pullin}. 

However, covariant canonical methods, to my knowledge, have never been tested in perturbation theory. Given the success of this approach for the classical theory on one hand, and skepticism regarding its implications in quantum theory on the other (see comments in Ashtekar et al. \cite{Ashtekar}), testing this idea on a toy model as a preliminary investigation merits attention. As a first step, we construct the quantum theory of the quartic oscillator via classical perturbation theory.  Section 3 of this paper carries the details of this construction. Further,as we shall discuss in the concluding section, the quartic oscillator model may provide signposts on how the operational procedure can be adapted to derive meaningful results for minisuperspace Bianchi models. Because of the high degree of symmetry in these models, they are effectively quantum mechanical problems, and provide the simplest nontrivial examples of dynamical spacetimes. There is also a large body of literature devoted to reducing the Einstein equations to equivalent Hamiltonian dynamical systems for these models. It may be possible to exploit these results if we know the steps in quantizing a Hamiltonian system via the proposed approach.
\section{Quantum mechanics with classical canonical data}
In the simplest route to canonical quantization one starts with an underlying classical structure and proceeds to define the Heisenberg algebra $\mathbb{H}$ of quantum observables via a transition from the Poisson bracket or symplectic 2-form \footnote{See Isham \cite{Isham2} for a review of algebraic quantization with the symplectic structure as a fundamental building block.} $\Omega=dp \wedge dq$ to the commutator $[\hat{q}, \hat{p}]$. As emphasized by \v Crnkovi\'c et. al \cite{Witten}, this process is not tied down to the seemingly non-covariant a priori choice of coordinates and momenta. Rather, the fundamental geometric structure on phase space, the symplectic 2-form, takes precedence. This follows from the observation that the presymplectic 1-form $\Theta =p_a dq^a$ is a completely covariant structure that can be read off from the boundary term in the variation of the action when the variation is restricted to the classical solutions to the equations of motion \cite{Wald, RobertWald}. In other words, when restricted to the classical solutions,
\begin{align}
\delta I_{\textrm{boundary}}&=\int dt \frac{d}{dt} ( p_a dq^a)=[\Theta] \bigg|_{t_i}^{t_f}{}\nonumber\\
\Omega:&=d \Theta
\end{align}
When the initial value formulation is well posed, classical solutions $(\bar{q}(t), \bar{p}(t))$ are generated uniquely from initial data $(q,p)$ by a canonical transformation $\mathbb{C}$
\begin{eqnarray}
\mathbb{C}:\big(q, p \big) \mapsto \bar{q}(t):=q \big(q, p; t \big);\quad
\bar{p}(t):=p \big(q, p; t \big)
\end{eqnarray}
By restricting the solutions to an arbitrary initial value surface $\Sigma$, this canonical transformation can then be inverted to represent data in terms of solutions, that is we can find expressions,  $q=q(\bar{q}, \bar{p}; t)$, $p=p(\bar{q}, \bar{p}; t)$.
Hence, the space of solutions of the equations of a motion is isomorphic to the space of initial data. However, each of the data points is also a point on the phase space. Thus the symplectic 2-form on phase space $\Omega$ can be pulled back to the space of data $\mathbb{S}$ to define a symplectic 2-form $\omega =dq \wedge dp =\mathbb{C}_* \Omega$ on the latter. Once this is done, quantization can be performed by a finding suitable self-adjoint operators 
$\hat{q}$ and $\hat{p}$, which generates the Heisenberg group $\mathbb{H}$. Wave functions $\psi(q)$ or $\phi(p)$ are defined for all times and are thus the wave functions of the Heisenberg picture of quantum mechanics. In passing let us also observe that, since data is independent of time, the total time derivative vanishes, that is  
\begin{align}
\dot{q}=\Big\{q, H \Big\}_{\textrm{PB}}+\frac{\partial q}{\partial t}=0
\end{align}
An ``evolving constant''$\mathcal{A}$, introduced by Rovelli \cite{Rovelli2} is a one parameter family of functions, parametrized by ``time'' $t$ that satisfies $\dot{\mathcal{A}}=0$. Hence from (2.4) we see that canonical data satisfies this criterion and hence are themselves evolving constants. 

When the full classical solutions are not known, we can re-formulate the problem of finding solutions in a perturbative sense outlined below.  To be concrete we will confine ourselves to a system with a Hamiltonian description. Let us assume that we have a system described by a Hamiltonian $H_0$ for which the full classical solutions are known.\footnote{In the proposal put forward by Bergmann et. al \cite{Bergmann}, it is assumed that the lowest order solution is the Minkowski metric, and the expansion is ``in terms of a deviation from this trivial constant solution''.} This in particular means that the observables of the unperturbed problem $(q_0, p_0)$ are known and that the unperturbed symplectic structure $\omega_{0} =dp_0\wedge dq_0$ is given. Assume then a small perturbation $H_1$, so that the full Hamiltonian is
\begin{align}
H=H_0+\epsilon H_1 
\end{align}
The idea is to use the full Hamiltonian to ``evolve'' the dynamical system, and in particular construct the observables of the problem described by $H$ 
in terms of the canonical data of the unperturbed problem. The canonical variables of the full problem are then obtained as a perturbation expansion defined as, 
\begin{align}
q&=q_0+\epsilon q_1+\epsilon^2 q_2 +....{}\nonumber\\
p&=p_0+\epsilon p_1+\epsilon^2 p_2+....
\end{align}
From the Hamilton's equations and the requirement that the observables of the full theory satisfy $\Big\{q, p\Big\}_{\textrm{PB}} =1$ at \emph{each order}, we  obtain solutions for  $(q_1, q_2)$ and $(p_1, p_2)$ and construct the solutions defined above at each order as functions of $\big(q_0, p_0 \big)$, and investigate the consequences of the perturbation on the zeroth order symplectic structure. The latter when promoted to a commutator algebra for the corresponding operators, encode the ``evolution'' of the unperturbed basis states that define the Hilbert space on which the quantum theory is defined.
We now turn to the details of application of this scheme to the quantum theory of the quartic oscillator. 
\section{Quantum theory of quartic oscillator via classical perturbation}
The complete Hamiltonian for the toy model is described by
\begin{eqnarray}
H=\frac{p^2}{2}+\frac{\omega^2 q^2}{2}+\frac{\epsilon q^4}{4}
\end{eqnarray}
The  basic canonical variables of the unperturbed harmonic oscillator problem will be denoted by $(q_0, p_0)$. The classical solutions of the unperturbed Harmonic oscillator is denoted as $\bar{q}(t):= q_0 \cos \omega t+\frac{p_0}{\omega} \sin \omega t$. In particular this means that for $t=0, \bar{q}=q_0$ and $\bar{p}=p_0$ represent oscillator data. Time dependent solutions will be denoted with an overbar through out the remainder of this text, while their restriction to the $t=0$ surface will be denoted without it.
The steps leading to the quantum theory can be outlined as follows:
\begin{enumerate}
\item  Hamilton's equations, in conjunction with the perturbation expansion 
\begin{align}
\bar{q}(t)&=\bar{q}_0 (t)+\epsilon \bar{q}_1( t)+\epsilon^2 \bar{q}_2 (t)+...{}\nonumber\\
\bar{p}(t)&=\bar{p}_0(t)+\epsilon \bar{p}_1(t)+\epsilon^2 \bar{p}_2 (t)+...
\end{align}
are used to construct the solutions as $\bar{q}=q(q_0, p_0; t), \bar{p}=p(q_0, p_0; t)$.
\item We impose $\{\bar{q}(t), \bar{p}(t)\}_{\textrm{PB}}=1$ to all orders in perturbation theory, and use this condition to obtain an expression for the Poisson bracket of the oscillator variables. We find that the Poisson algebra is preserved under time evolution, hence we work with physical quantities defined at $t=0$. The perturbation manifests itself as a deformation of the classical Poisson algebra of $(q_0, p_0)$. The Hamiltonian (3.6) is expanded to each order in perturbation theory with the aid of the solutions. The equations of motion ensure that it is an evolving constant in the sense of section 2. 
\item  We define a set of new canonical variables, expressed as functions of $(q_0, p_0)$ at each order in perturbation, which satisfies a classical canonical algebra. The introduction of this canonical pair is motivated by the adjustment of the deformation in the classical Poisson algebra of the oscillator variables (see section 3.1 for an explanation on the meaning of the term adjustment used here). This enables a transition to a canonical pair of operators that generates a Heisenberg group $\mathbb{H}$ at each order of perturbation. This is needed to construct the Hilbert space of states and to facilitate operator ordering of the Hamiltonian.
\item We promote the latter variables with the choice of Weyl ordering to hermitian operators acting on the Hilbert space $\mathcal{H}_{\textrm{osc}}$ of the unperturbed oscillator states, and construct a Hamiltonian operator $\hat{H}$ whose action on these states gives the quantized spectra at each order of perturbation.
\end{enumerate}

Proceeding in step 1, Hamilton's equations lead to the differential equations
\begin{align}
\dot{\bar{p}}=-\omega^2 \bar{q}+\epsilon \bar{q}^3;\quad \dot{\bar{q}}=\bar{p}{}\nonumber\\
\Rightarrow \ddot{\bar{q}}+\omega^2 \bar{q}+\epsilon \bar{q}^3=0
\end{align}
These must be satisfied at each order in $\epsilon$, and in particular yield the equations of motion for 
for $\bar{q}_1$, $\bar{p}_1$, $\bar{q}_2$, $\bar{p}_2$ respectively as
\begin{align} 
\ddot{\bar{q}}_{1}+{\omega}^2\bar{q}_{1}&=-{\bar{q}_0}^3 (t){}\nonumber\\
\bar{p}_1&=\dot{\bar{q}}_1{}\nonumber\\
\ddot{\bar{q}}_{2}+{\omega}^2\bar{q}_{2}&=-3\bar{q_0}^2\bar{q}_1 (t){}\nonumber\\
\bar{p}_2&=\dot{\bar{q}}_2 
\end{align}
where $\bar{q} (t) =q_0 \cos \omega t+\frac{p_0}{\omega} \sin \omega t$. 
The solutions for $\bar{q}_1 (t)$ and $\bar{p}_1 (t)$, are then given by
\begin{align}
\bar{q}_1 (t)&= \frac{\zeta_1 t \textrm{sin}\omega t}{2\omega}-\frac{\zeta_2 t \textrm{cos}
\omega t}{2\omega}-\frac{\zeta_3 \textrm{sin}3\omega t}{8 {\omega}^2} -\frac{\zeta_4 \textrm{cos}3\omega t}{8 {\omega}^2}{}\nonumber\\
\bar{p}_1(t)&=\frac{\zeta_1}{2\omega} \big(t\omega \cos \omega t+\sin \omega t \big)-\frac{\zeta_2}{2 \omega} \big(-t \omega \sin \omega t+\cos \omega t \big)-\frac{3 \zeta _3}{8 \omega} \cos 3\omega t +\frac{3 \zeta _4}{8 \omega} \sin 3 \omega t
\end{align}
where the set of coefficients $\zeta_1$, $\zeta_2$, $\zeta_3$, and $\zeta_4$ are given by
\begin{align}
\zeta_1 (q_0, p_0)&=-\frac{3}{4} \bigg(q_0^3+\frac{q_0 p_0^2}{4\omega^2} \bigg){}\nonumber\\
\zeta_{2} (q_0, p_0)&=-\frac{3}{4} \bigg(p_0^3 +\frac{q_0^2 p_0}{\omega} \bigg){}\nonumber\\
\zeta_{3} (q_0, p_0)&=\frac{1}{4} \bigg(\frac{p_0^3}{\omega^3} -\frac{3 q_0^2 p_0}{4 \omega} \bigg){}\nonumber\\
\zeta_{4} (q_0, p_0)&=\frac{1}{4} \bigg( \frac{3q_0 p_0^2}{4 \omega^2} -q_0^3 \bigg)
\end{align}
As a consequence of the equations of motion, $\frac{d H}{dt}=0$. Additionally, a tedious but straightforward exercise shows that $\{\bar{q}(t), \bar{p}(t) \}_{\textrm{PB}}=\{q, p \}_{\textrm{PB}}$. These observations imply that ``time'' $t$ which appears in the solutions (3.5) plays the role of a parameter and that there is no physical significance to a particular instant in $t$. This justifies working with values of $(q, p)$ at $t=0$, without any loss of generality. 
Thus we will work at $0(\epsilon)$, with the solutions
\begin{align}
\bar{q}_1 (0)&=q_1=\frac{q_0^3}{32\omega^2}-\frac{3q_0p_0^2}{32\omega^4}{}\nonumber\\
\bar{p}_1(0)&=p_1=\frac{9 p_0^3}{32\omega^4}+\frac{21q_0^2p_0 }{32\omega^2}
\end{align}
and the corresponding ones at $O(\epsilon^2)$, which are given by
\begin{align}
q_2&=-\frac{5 q_0^5}{256\omega^4} +\frac{8q_0^3p_0^2 }{256\omega^6}+\frac{17q_0p_0^4 }{256\omega^8}{}\nonumber\\
p_2&=-\frac{56 q_0^2p_0^3}{256\omega^6}-\frac{29q_0^4p_0 }{256\omega^4}-\frac{7p_0^5 }{256\omega^8}
\end{align}

We will proceed first with computations at the first order in perturbation theory. For notational simplicity, we introduce two parameters, $\beta=\frac{3}{2\omega^4}$ and  $\kappa=\beta\epsilon\hbar\omega$. Henceforth, $H_0 =\frac{p_0^2}{2}+\frac{\omega^2 q_0^2}{2}$. 
Now using $\{q, p\}_{\textrm{PB}} =1$ and the solutions (3.7) we obtain for the Poisson bracket of the original variables
\begin{align}
\Big\{q_0, p_0 \Big \}_{\textrm{PB}}&=\Big(1-\beta \epsilon H_0 \Big)
\end{align}
Thus the perturbation leads to a deformation of the Poisson algebra of the unperturbed observables $\big(q_0, p_0 \big)$. This in particular implies that the Poisson bracket of arbitrary phase space functions $f(q_0, p_0)$ and $g(q_0, p_0)$ is given by 
\begin{align}
\Big\{f, g \Big\}_{\textrm{PB}}&=\Big(1-\beta \epsilon H_0 \Big){\Big\{f, g \Big\}}^{(0)}_\textrm{PB}
\end{align} 
where the superscript on the Poisson bracket implies zeroth order. With the help of the solutions the Hamiltonian at first order can be written as
\begin{align}
H(q_0, p_0)&=\frac{(p_0+\epsilon p_1)^{2}}{2}+\frac{\omega^2 (q_0+\epsilon q_1)^{2}}{2}+\frac{\epsilon q_0^4}{4}{}\nonumber\\
&=H_0+\frac{3 \beta \epsilon}{4} H_0^2
\end{align}
and at second order from (3.7) and (3.8)
\begin{eqnarray}
H(q_0, p_0)=H_0+\frac{3 \beta \epsilon}{4} H_0^2+\frac{25 \beta^2 \epsilon^2}{576} H_0^3
\end{eqnarray}

We now turn to the quantum theory of this system. 
\subsection{Symplectic structure, quantization and spectrum}
In order to obtain the spectrum of energies for the Hamiltonian (3.11) and (3.12), we need to promote the variables $(q_0, p_0)$ and the Hamiltonian $H$ to hermitian operators on the Hilbert space $\mathcal{H}_{\textrm{osc}}$ of the harmonic oscillator. The key to this quantization is the deformation of the classical Poisson bracket algebra (3.9), which when promoted to a commutator via the correspondence, $[, ]\rightarrow i\hbar \{, \}$ yields
\begin{eqnarray}
\big[\hat{q}_0, \hat{p}_0 \big] =i \hbar \big(1-\beta \epsilon \hat{H}_0  \big)
\end{eqnarray}
Observe that ---
\begin{enumerate}
\item  The usual Heisenberg group $\mathbb{H}$ in quantum mechanics is generated by the algebra 
$[\hat{q}, \hat{p}]=i\hbar$. Owing to the deformation of the classical Poisson algebra, $\hat{q}_0$ and $\hat{p}_0$ do not have this property. This renders the algebraic problem of operator ordering, such as Weyl ordering for example, unusually complex.
\item  The spirit of the present proposal is to use the Hilbert space of states defined by the unperturbed harmonic oscillator to construct the quantum theory of the perturbed problem at each order. Note, however, that covariant canonical quantization is the Heisenberg picture of quantum theory. It is well known that basis states of the Heisenberg picture evolve, and in this case, this evolution is imprinted as a deformation of the algebra. Hence, we need to to define a canonical pair generating a Heisenberg group at each order and to construct  basis states $|\widetilde{osc}> \in \mathcal{H}_{\textrm{osc}}$, on which we can write the action of the Hamiltonian operator. 
\end{enumerate}
These observations motivate the transformation to a very simple canonical pair $(\tilde{q}, \tilde{p})$ that can be used to quantize the system at each order in perturbation theory. This transformation is what we referred to as an ``adjustment'' at the beginning of section 3. The new variables will be referred to throughout the remainder of this paper as a tilded-representation. 
The choice of this canonical pair is naturally motivated by the structure of the deformation of the classical Poisson algebra of the unperturbed oscillator variables.  
At $0(\epsilon)$ the new variables are  
\begin{align}
\tilde{q} (q_0, p_0):&=q_0 \bigg(1+\frac{\beta \epsilon}{4} H_0\bigg) {}\nonumber\\
\tilde{p} (q_0, p_0):&=p_0 \bigg(1+\frac{\beta \epsilon}{4}H_0 \bigg)
\end{align}
This pair satisfies
\begin{equation}
\Big\{\tilde{q}, \tilde{p} \Big\}_{\textrm{PB}}=1
\end{equation}
The classical Hamiltonian (3.16) now becomes
\begin{eqnarray}
H(\tilde{q}, \tilde{p}):=\tilde{H}_0+\frac{\beta \epsilon}{4} \tilde{H}_0^2
\end{eqnarray}
where
\begin{eqnarray}
\tilde{H}_0 =\frac{\tilde{p}^2}{2}+\frac{\omega^2 \tilde{q}^2}{2}= H_0 -\frac{\beta \epsilon}{2} H_0^2 +O(\epsilon^2)
\end{eqnarray}
We define a quantum algebra by Weyl ordering the pair (3.14),  
\begin{align}
\tilde{q}\mapsto [\tilde{q}]_{\mathcal{W}}&=\hat{\tilde{q}}=\bigg(\hat{q}_0+\frac{\beta\epsilon}{8}\big[\hat{q}_0, \hat{H}_0 \big]_{+}\bigg){}\nonumber\\
\tilde{p}\mapsto [\tilde{p}]_{\mathcal{W}}&=\hat{\tilde{p}}=\bigg(\hat{p}_0+\frac{\beta\epsilon}{8}\big[\hat{p}_0, \hat{H}_0 \big]_{+}\bigg)
\end{align}
To $O(\epsilon)$, therefore, we have a canonically conjugate pair which are functions of the oscillator data and which generates a Heisenberg group,
\begin{equation}
\Big[\hat{\tilde{q}}, \hat{\tilde{p}} \Big] =i \hbar
\end{equation}

A word in passing as to the meaning of Weyl ordering used in the above definition would be appropriate. Owing to the deformation in the commutator algebra eq.\ (3.9),  we would expect a $O(\epsilon)$ modification of the standard Weyl ordering \cite{mccoy}. However, this does not contribute at $O(\epsilon)$ to the definitions (3.18), which incorporates the zeroth Weyl-ordered expression. Once these tilded-variables are defined, operator ordering to obtain $0(\epsilon^2)$  expressions for operators corresponding to polynomial functions of $(q_0, p_0)$ does  not pose a problem, since such functions can be re-expressed in the tilded representation by inverting the relations (3.14), using eq.\ (3.16) where appropriate. By construction the operator of the tilded-representation generate a Heisenberg group, and hence usual Weyl ordering applies to functions of the form $f(\tilde{q}^m \tilde{p}^r)$. 

 We now define a basis in terms of the operators
\begin{align}
\hat{\tilde{a}}_\omega&=\sqrt{\frac{\omega}{2\hbar}}\Big(\hat{\tilde{q}}+\frac{i}{\omega}\hat{\tilde{p}}\Big){}\nonumber\\
{\hat{\tilde{a}}^\dagger}_\omega&=\sqrt{\frac{\omega}{2\hbar}}\Big(\hat{\tilde{q}}-\frac{i}{\omega}\hat{\tilde{p}}\Big){}\nonumber\\
\hat{\tilde{N}}&={\hat{\tilde{a}}^\dagger}_\omega \hat{\tilde{a}}_\omega=\frac{\hat{\tilde{H}}_0}{\hbar \omega} -\frac{1}{2}
\end{align}
where $\hat{\tilde{H}}_0=\frac{\hat{\tilde{p}}^2}{2} +\frac{\omega^2 \hat{\tilde{q}}^2}{2}$. Owing to eq.\-(3.19), $[\hat{\tilde{a}}_\omega,{\hat{\tilde{a}}^\dagger}_\omega]=1$ and hence $\hat{\tilde{a}}_\omega$ and ${\hat{\tilde{a}}^\dagger}_\omega$ satisfy the algebra of the usual creation-annihilation operators. As a consequence we can define states $|\tilde{n}>$ that satisfy $\hat{\tilde{N}}|\tilde{n}>=\tilde{n} |\tilde{n}>$ with $\tilde{n} \in {\mathbb{Z}}^{+}$.  A straightforward calculation shows that the Hamiltonian (3.11) can be recast in terms of the Heisenberg pair (3.18), which upon yields the quantum Hamiltonian at $O(\epsilon)$,  
\begin{eqnarray}
\hat{H}:=\hat{\tilde{H}}_0+\frac{\beta \epsilon}{4}\hat{\tilde{H}}_0^2+\frac{\kappa \hbar \omega}{16}\hat{\mathbb{I}}
\end{eqnarray}
Given the definitions of eq.\ (3.20), the spectrum of $\hat{H}$ is given by (reinstating $\beta$ and $\kappa$)
\begin{eqnarray}
\hat{H} \Big|\tilde{n} \Big> =\bigg[\Big(\tilde{n}+\frac{1}{2}\Big)\hbar \omega+\frac{3 \epsilon \hbar ^2}{8\omega^2}\Big(\tilde{n}^2+\tilde{n}+\frac{1}{2}\Big)\bigg] \Big|\tilde{n}\Big>
\end{eqnarray} 
The energy eigenvalues are in exact agreement with the results of Rayleigh-Schr\"odinger perturbation theory \cite{Flugge}, implying thereby that our method reproduces the known result for the spectrum of the quartic oscillator in perturbation theory at this order. 

Computations to second order are straightforward, although algebraically more complex owing to the fact that the effect of the first order perturbation have to be taken into account at this order.  First, the solutions (3.7) and (3.8)  and eq.\ (3.10)  imply the following deformation of the Poisson bracket of the unperturbed variables at second order:
\begin{eqnarray}
{\Big\{q_0, p_0 \Big \}}_{\textrm{PB}} = \Big(1 -\beta\epsilon H_0 +\frac{81 \beta^2 \epsilon^2}{64} H_0^2 \Big)
\end{eqnarray}
We observe already at this order that there is a emergent pattern in the way the perturbation appears as a deformation of the classical canonical algebra, the deformation being proportional to the unperturbed Hamiltonian raised to the power of the order of the perturbation. Given  eq.\ (3.23) we can now define the tilded- pair at second order in analogy with (3.14) as
\begin{align}
\tilde{q}:&=q_0 \bigg(1+\frac{\beta \epsilon}{4} H_0 -\frac{29 \beta^2 \epsilon^2}{384} H_0^2 \bigg){}\nonumber\\
\tilde{p}:&=p_0 \bigg(1+\frac{\beta \epsilon}{4}H_0 -\frac{29 \beta^2 \epsilon^2}{384}H_0^2 \bigg)
\end{align}
We proceed with steps similar to those at first order. First, eq.\ (3.12) gives the full classical Hamiltonian at $O(\epsilon^2)$ , which when transformed to the tilded-representation becomes
\begin{eqnarray}
H = \tilde{H}_0+\frac{\beta \epsilon}{4}\tilde{H}_0^2 -\frac{17 \beta^2 \epsilon^2}{144}\tilde{H}_0^3
\end{eqnarray}
Weyl ordering leads to the corresponding quantum Hamiltonian to $O(\epsilon^2)$,  
\begin{eqnarray}
\hat{H}:= \hat{\tilde{H}}_0+\Big[\frac{\beta \epsilon}{4}\hat{\tilde{H}}_0^2 +\frac{\kappa\hbar\omega}{16}\hat{\mathbb{I}}\Big] -\Big[\frac{17 \beta^2 \epsilon^2}{144} \hat{\tilde{H}}_0^3+\frac{67\kappa^2}{576} \hat{\tilde{H}}_0 \Big]
\end{eqnarray}
which satisfies the eigenvalue equation,
\begin{align}
\hat{H}\Big|\tilde{n} \Big>= \bigg[\Big(\tilde{n}+\frac{1}{2}\Big)\hbar \omega+\frac{3 \epsilon \hbar ^2}{8\omega^2}\Big(\tilde{n}^2+\tilde{n}+\frac{1}{2}\Big)-\frac{\epsilon^2 \hbar^3}{128 \omega^5}\Big(2\tilde{n}+1\Big)\Big(17 \tilde{n}^2 +17 \tilde{n}+21\big)\bigg]\Big|\tilde{n} \Big>
\end{align}
giving the quantized spectrum to second order that is once again in agreement with the standard quantum theory result.

Having thus carried out computations to second order in perturbation theory, we are now in a position to suggest an algorithmic procedure for constructing a quantum theory to higher orders. 

To do so, observe that at first order the perturbation manifests as the deformation of the Poisson algebra of unperturbed variables $(q_0, p_0)$ given by eq.\ (3.9), which leads to the definition of the new variables given by eq.\ (3.14) which satisfy a canonical algebra at this order. Next, owing to eq.\ (3.10) and (3.23), if one now calculates the Poisson bracket of these new variables of first order, to include all terms of $0(\epsilon^2)$, one finds
\begin{align}
\Big\{\tilde{q}, \tilde{p} \Big\}_{\textrm{PB}}=1+\frac{29 \beta^2 \epsilon^2}{64} H_0^2
\end{align}
This deformation in the Poisson algebra of the new variables of first order leads to the definitions eq.\ (3.24) of the new variables at second order, where with a slight abuse of notation we rename variables that generate the Heisenberg group at each order to be $(\tilde{q}, \tilde{p})$. This suggests that for this model, we can iteratively define new variables that can be used to construct a quantum theory at each order in perturbation theory in the following way: 

If at any given order of perturbation $n$, we have a deformation of the commutative Poisson algebra, $\{Q, P \}_{\textrm{PB}} =(1- \alpha_1  \Lambda)$ of generic conjugate pairs $Q=Q(q_0, p_0)$ and $P=P(q_0, p_0)$, such that $\Lambda(q_0, p_0)$ is a homogeneous function of degree $n$, and $\alpha_1 \sim O(\epsilon^n)$ is a constant--- then, ``new variables'', 
\begin{align}
\tilde{q}&=Q\bigg(1+\frac{\alpha_1}{n+2}\Lambda \bigg){}\nonumber\\
\tilde{p}&=P\bigg(1+\frac{\alpha_1}{n+2} \Lambda \bigg)
\end{align}
satisfy a canonical algebra at that order. The corresponding ordered pair $(\hat{\tilde{q}}, \hat{\tilde{p}})$ generate a Heisenberg group. To see this note that---
\begin{align}
\bigg\{\tilde{q}, \tilde{p} \bigg\}_\textrm{PB}&=\bigg\{Q\Big(1+\frac{\alpha_1}{n+2}\Lambda \Big),P\Big(1+\frac{\alpha_1}{n+2} \Lambda \Big)\bigg\}_\textrm{Pb}{}\nonumber\\
&\sim \bigg\{Q, P \bigg \}_\textrm{PB}+\frac{\alpha_1}{n+2}{\bigg\{\Lambda Q, P \bigg\}}^{(0)}_\textrm{PB}+\frac{\alpha_1}{n+2}{\bigg\{Q, \Lambda P\bigg\}}^{(0)}_\textrm{PB}{}\nonumber\\
&=\bigg(1-\alpha_1 \Lambda \bigg)+\frac{\alpha_1}{n+2} (n+2) \Lambda{}\nonumber\\
&=1
\end{align}
where we have use the fact that at zeroth order $Q=q_0$ and $P=p_0$ and have exploited Euler's theorem on homogeneous functions, which in particular means that $q_0\frac{\partial \Lambda}{\partial q_0}+p_0\frac{\partial \Lambda}{\partial p_0}=n \Lambda$.
  
The conclusion is that at each order, with the help of the perturbative solutions a corresponding quantum theory can be constructed and that with a choice of ordering (in this case Weyl ordering) we can pass at each level of perturbation theory from the classical evolving constants to their corresponding quantum counterparts.

We now turn to a a discussion of the relevance of the results for quantum gravity.

\section{Conclusion and future outlook}
The results of the quartic oscillator model demonstrates  through an explicit example that it is possible to extend covariant canonical quantization to include perturbation theory.  Thus, I have demonstrated that where we do not a priori know the full classical solution, we can resort to a two-step procedure to find the corresponding quantum observables, namely 
\begin{enumerate}
\item Construct classical solutions via Hamilton's equations to perturbatively generate the solution space $\mathbb{S}$ at each order.
\item Quantize the solution space thus generated using as the Hilbert space the time evolved basis states of the unperturbed problem. 
\end{enumerate}
The result also addresses an issue raised in the past by Hajicek \cite{hajicek} regarding computability of  ``evolving constants'', while at the same time providing a concrete example of how to implement of Bergmann's proposal for a covariant approximation scheme in quantum theory. An immediate application may be the quantum theory of minisuperspace models in general relativity which by virtue of their small number of degrees of freedom are quite tractable, yet physically relevant \cite{Kuchar}. Recently such models have attracted a lot of attention as test beds of quantization techniques \cite{husain, bombelli, bojowald, bojowald2} . For instance, a problem of great physical interest is the study of the BKL behavior near cosmological singularities in quantum theory. Furthermore, dynamical system approaches where the Einstein equations can be reduced to a set of autonomous differential equations exist for such models \cite{coley, ellis}. This opens up immediate possibilities of exploiting Hamiltonian techniques \cite{Waldron, Nicolai} to study the quantum theory of such systems. If this bears fruit, then the next step would be to apply covariant canonical quantization to midisuperspaces, such as those of Gowdy models, whose classical data have recently been studied by several authors \cite{moncrief,beverly,D,quevedo,garfinkle}. These questions are currently being investigated and will be reported in future publications.

\textbf{Acknowledgments}
I would like to thank Steven Carlip for suggesting the problem, as well as providing many insightful discussions. Thanks are due to David Mattingly for useful comments and to Mike Chu for pointing out certain relevant issues. I would also like to thank the College of Letters and Science, UC Davis, for a summer research fellowship. This work was funded under 
DOE grant DE-FG02-91ER40674.

\end{document}